%% file: Paulucci.tex
\documentclass[12pt]{article}
\usepackage{epsfig}

\input econfmacros-csqcd3.tex
\def\Title#1{\begin{center} {\Large {\bf #1} } \end{center}}

\begin{document}

\Title{Superconducting phases of strange quark matter in the NJL model}

\bigskip\bigskip


\begin{raggedright}

{\it Laura Paulucci\index{Paulucci, L.}\\
Universidade Federal do ABC\\
Rua Santa Ad\'elia, 166, 09210-170 Santo Andr\'e, SP, Brazil \\
{\tt Email: laura.paulucci@ufabc.edu.br}} \\
{\it J. E. Horvath\index{Horvath, J. E.}\\
Instituto de Astronomia, Geof\'\i sica e Ci\^encias Atmosf\'ericas, Universidade de S\~ao Paulo\\
Rua do Mat\~ao 1226, 05508-900 S\~ao Paulo SP, Brazil}\\
{\it E. J. Ferrer\index{Ferrer, E. J.} and V. de la Incera\index{de la Incera, V.}\\
Department of Physics, University of Texas at El Paso\\
El Paso, TX 79968, USA}

\bigskip\bigskip
\end{raggedright}

\section{Introduction}

After the nuclear burning of its composing elements, stars born with masses greater than $\sim 8 M_{\odot}$ may end their evolutionary path as a compact object with roughly one solar mass compressed to a radius of about 10 km, known as a neutron star. The pressure in its interior is extremely large, with matter subjected to conditions not attainable yet (and perhaps never to be) in laboratories of high density and relatively low temperatures. The behavior of matter under these circumstances is still unknown with a large range of possibilities, from superfluid and superconductor nucleons to condensate of pions, kaons, and even the complete dissociation of hadrons into their basic constituents, quarks (see, for example, ~\cite{Fridolin}).

Even if conditions for a deconfinement phase transition are achieved in the interior of neutron stars, details on how the ``soup'' of quarks and gluons would behave are still blurred by the lack of a proper model to work in a non-perturbative scale. The most widely used models to study quark matter are the MIT bag model and the Nambu-Jona-Lasinio theory. In order to find out if quark matter may be a component of compact stars' interior, we must use these models to predict their properties and compare with observations.

It was in the late 1970's~\cite{dinos} that the possibility that strange quark matter, matter composed of quarks u, d and s, could be the lower energy state of high density and low temperature baryonic matter was raised. This stability could be achieved by the introduction of a third Fermi sea corresponding to the strange quark, if its mass would not be high enough to provide an energetic advantage for the weak decay of quarks u and d.

It was shown by Fahri and Jaffe~\cite{Farhi} that working within the MIT bag model there was room for absolute stability to hold, depending on some parameters (still barely known with precision) such as the strange quark mass, the bag constant and strong coupling constant. If this is the case, once conditions for the phase transition in the interior of neutron stars were achieved, there would follow a full conversion of the star, rendering a strange star \cite{SS}. If strange quark matter was only metastable, there could be the formation of hybrid stars, in which only their core would be composed of quark matter.

Latter on, it has been put forward the idea that pairing between quarks (like the phenomenon of electronic superconductivity) could lower the energy per baryon number of the matter even further, increasing the window of stability~\cite{Pairing, German}. It would be a natural condition for quark matter, since the interaction between quarks has an attractive channel. Since these developments, much work has been done in order to establish the phase diagram of baryonic matter in the low temperature range. As can be seen in~\cite{Buballa}, for example, the subject is still under debate and there are still open questions.

The most widely used constraint on the composition of a neutron star is the maximum mass it can support against gravitational collapse. A recent and very precise measurement by Demorest et al.~\cite{Demorest} of PSR J1614-2230 indicates a mass of $1.97\pm 0.04 M_{\odot}$ for this pulsar. Whatever the composition of these stars, it must be able to provide enough pressure to support at least two solar masses.

We are going to discuss here the possible existence of quark matter in the interior of these compact objects, presenting the equation of state for superconducting quark matter with pairing within the NJL model, comparing with results obtained when using the MIT bag model. We will also briefly discuss the influence of magnetic fields in the mass-radius relation.

\section{Color-Flavor-Locked Matter}

We consider the three-flavor Nambu-Jona-Lasinio theory and neglect all quark masses. This ensures color and electrical neutrality and there is only one chemical potential, the baryonic one $\mu$. The thermodynamic potential of the color-flavor-locked (CFL) phase, when all quarks are paired, is

\begin{equation}
\Omega_{CFL} =-\frac{1}{4\pi^2}\int_0^\infty dp p^2 e^{-p^2/\Lambda^2}[16|
\varepsilon|+16|\overline{\varepsilon}|]-\frac{1}{4\pi^2}\int_0^\infty
dp p^2 e^{-p^2/\Lambda^2}[2|\varepsilon'|+2|\overline{\varepsilon'}|]+
\frac{3\Delta^2}{G} \label{OmegaCFL}
\end{equation}
where $\Lambda$ is introduced to give a smooth cutoff with value 1 GeV and

\begin{eqnarray*}\label{6}
\varepsilon=\pm \sqrt{(p-\mu)^2+\Delta^2}, \nonumber \,\,\,\,\,\,\,
\overline{\varepsilon}=\pm \sqrt{(p+\mu)^2+\Delta^2},\nonumber
\\
\varepsilon'=\pm \sqrt{(p-\mu)^2+4\Delta^2}, \,\,\,\,\,\,\,
 \overline{\varepsilon'}=\pm \sqrt{(p+\mu)^2+4\Delta^2},\nonumber
\end{eqnarray*}
are the quasiparticles' dispersion relations.

The value of the pairing gap $\Delta$ is determined by solving the gap equation

\begin{equation}
\frac{\partial \Omega}{\partial \Delta}=0
\end{equation}
and the equations of state are given by

\begin{eqnarray}
\epsilon &=& \Omega - \mu \frac{\partial \Omega}{\partial \mu} \\
p &=& -\Omega
\end{eqnarray}

\begin{figure}[htb]
\begin{center}
\epsfig{file=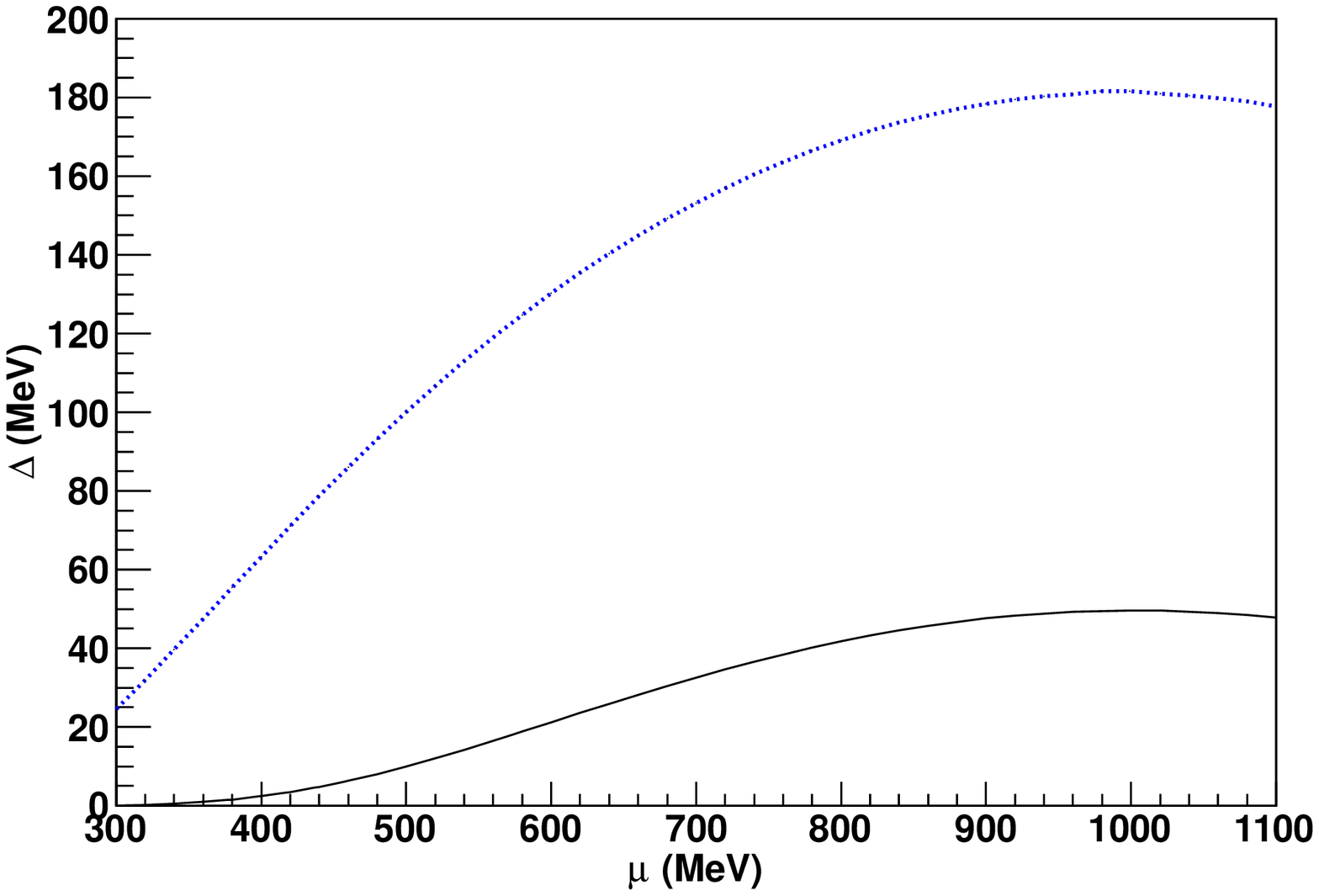,width=0.49\textwidth}
\epsfig{file=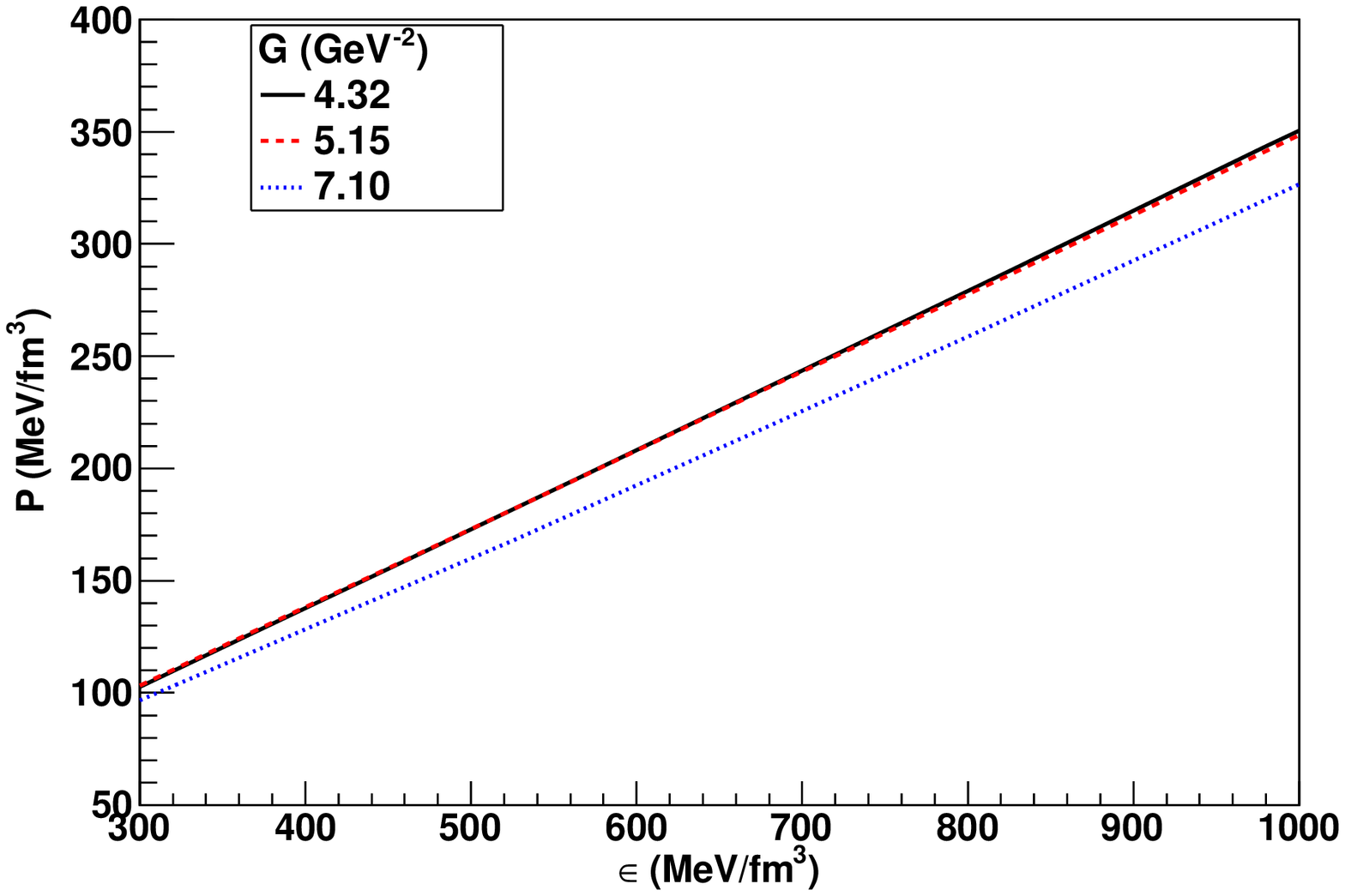,width=0.49\textwidth}
\caption{Dependence of the gap parameter $\Delta$ with the chemical potential $\mu$ (left) and equation of state (right) for CFL matter in the NJL approach for different values of $G$, as indicated.}
\label{NJL}
\end{center}
\end{figure}

Using the values of G=4.32 GeV$^{-2}$, G=5.15 GeV$^{-2}$, and G=7.10 GeV$^{-2}$,
it is obtained, at $\mu=500$ MeV, the gaps $\Delta=10$, 25, and 100 MeV, respectively.
The dependence of $\Delta$ with chemical potential is given in the left panel of Fig.~\ref{NJL}, and the equation of state in the region of
interest for strange stars is shown in the right panel of Fig.~\ref{NJL}.

\subsection{Comparing models}

The behavior seen for the equation of state in Fig.~\ref{NJL} is in contrast with the one seen when considering CFL matter in the MIT bag model. The thermodynamic potential for the system up to order
$\Delta^2$ considering the model proposed in \cite{toy} is

\begin{equation}
\Omega_{CFL}=\sum_i\Omega_i-\frac{3}{\pi^2}\Delta^2\mu^2+B \label{MIT}
\end{equation}
\\
where the extra term dependent on $\Delta$ represents the binding energy of the diquark
condensate. The term $\Omega_{free}=\sum_i\Omega_i$ represents a fictitious non-paired
state in which all quarks have a common Fermi momentum. In (\ref{MIT}), B is the bag constant, which incorporates the confinement effects.

In this way, since $\Delta$ is a parameter fixed by hand in the calculations, the EoS
one obtains by changing its value leads to the conclusion that the higher the gap, the stiffer the EoS, i. e., the higher the maximum allowed mass for the strange star.
Note, however, that this is not the case for the NJL calculations. When a higher
value of $G$ is used, the EoS {\it softens} in the region of interest.

For calculations using (\ref{MIT}) one can increase the maximum allowed mass
for strange stars by using a higher gap but not with NJL, where the gap has to be found self-consistently through the gap equation. Although being a
simplified model, the MIT result seems intuitive, but we have to have in mind that the $\Delta$ values are not arbitrary, as shown in the NJL self-consistent approach. It would be interesting to
understand the reasons why it is not the case with the NJL.

\begin{figure}[htb]
\begin{center}
\epsfig{file=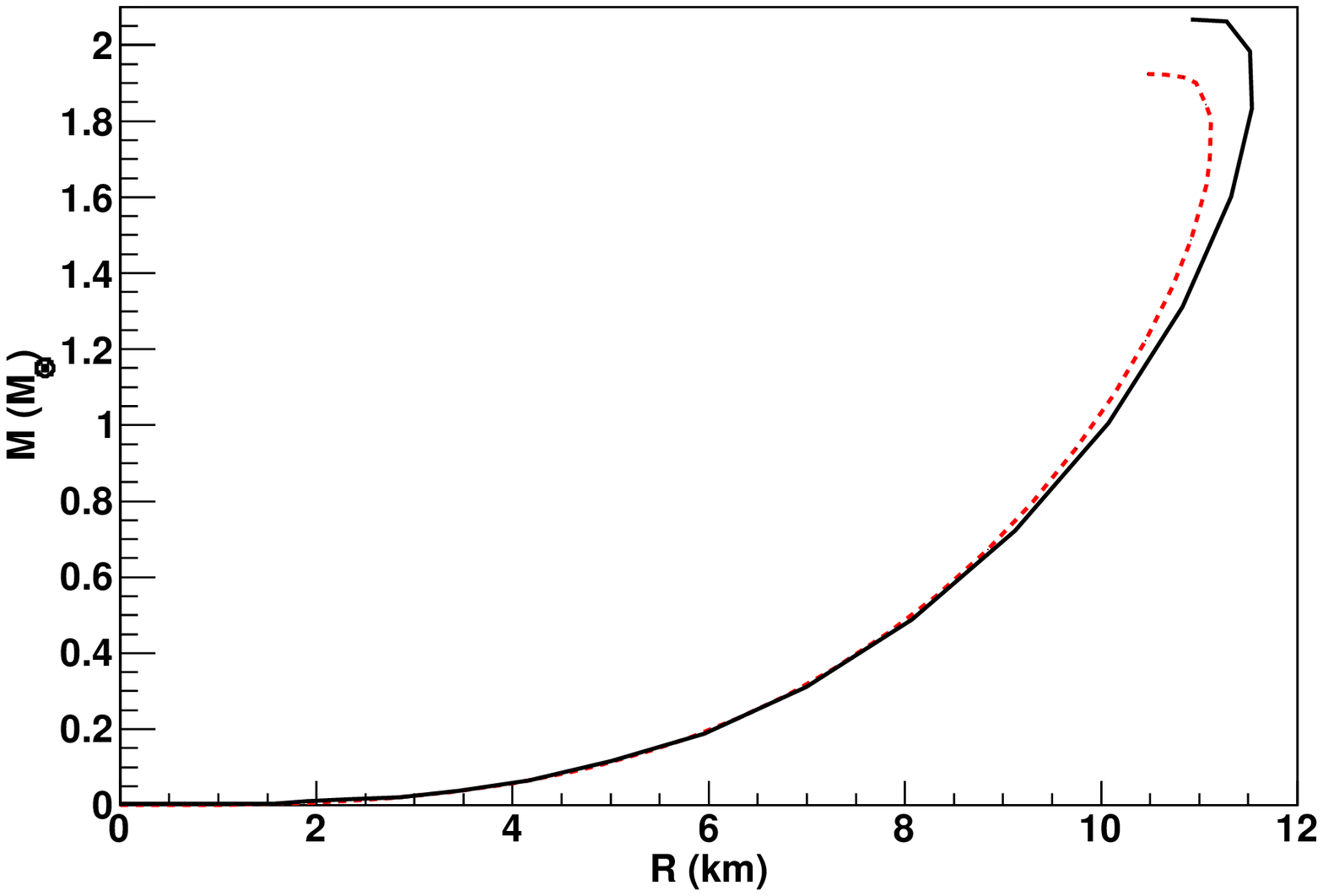,width=0.49\textwidth}
\epsfig{file=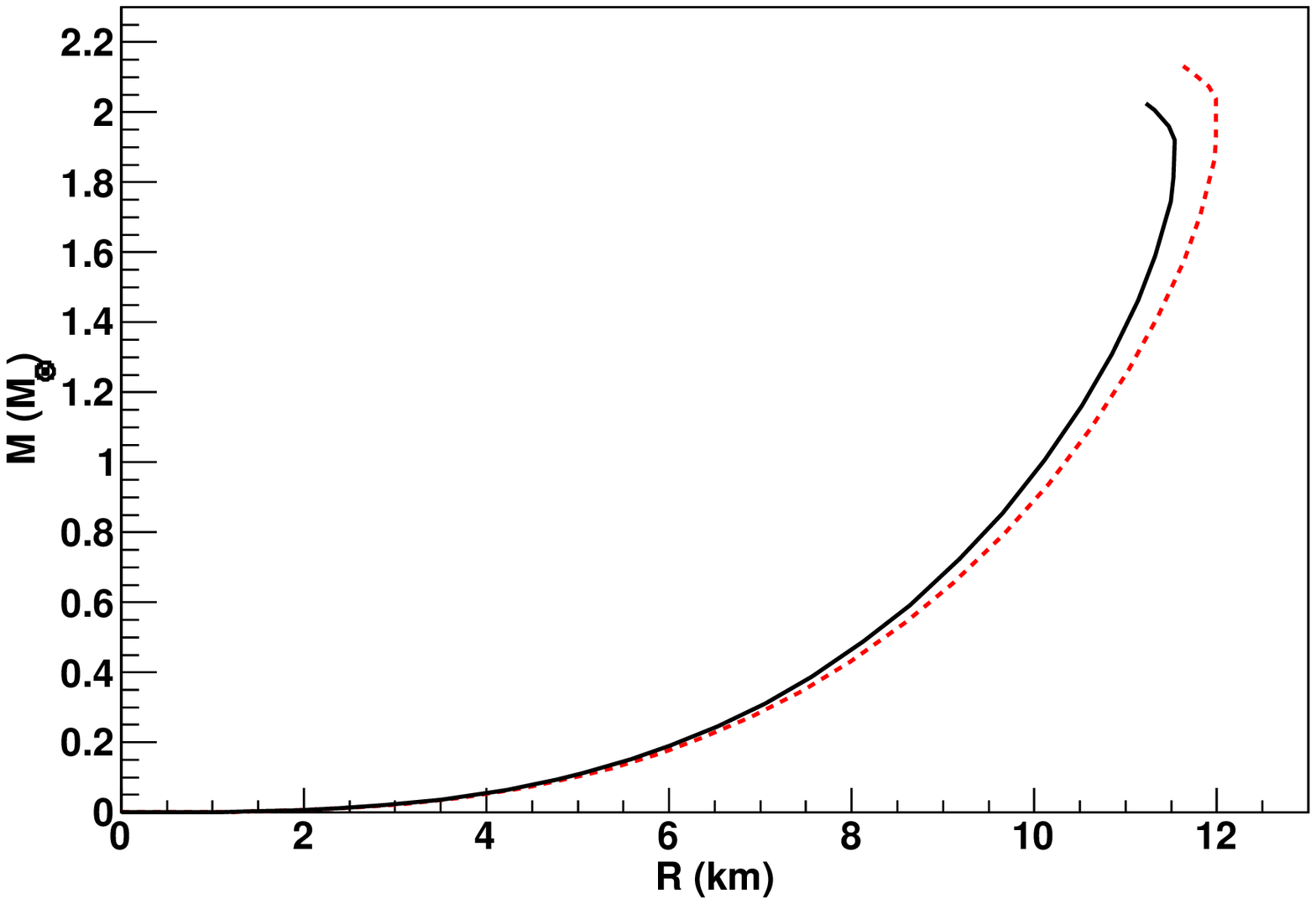,width=0.49\textwidth}
\caption{Mass-radius relation for CFL matter using $B=58$ MeV/fm$^3$ (also added to equation \ref{OmegaCFL}). On the left panel, NJL theory, where the full line is for $G$=4.32 GeV$^{-2}$ and dashed line for $G$=7.10 GeV$^{-2}$. On the right, MIT bag model with $\Delta=10$ and 100 MeV for full and dashed line, respectively.}
\label{MR}
\end{center}
\end{figure}

\subsection{The influence of magnetic fields}

The magnetic field on the surface of a neutron star ranges from B = $1.7 \times 10^8$G to $2.1 \times 10^{13}$G, and may reach B $\sim 10^{14 - 15}$G when considering magnetars. A crude estimate of the interior field can be given by applying the equipartition theorem, giving values up to $\sim 10^{19-20}$G \cite{Efrain}.

Considering the influence of an uniform magnetic field, the CFL superconductor can
be modeled by the NJL theory, the MCFL phase, where the effects of confinement are incorporated by introducing a bag constant $B$ in the thermodynamical potential of the phase, please refer to \cite{Paulucci} and references therein for more details.

When calculating the pressure and energy density of the MCFL phase \cite{MCFL} the pure magnetic field contribution and the magnetization of the matter must be added, resulting in an anisotropy in the pressure:

\begin{equation} \label{EOS-MCFL}
\epsilon_{MCFL}=\Omega_{H}-\mu \frac{\partial \Omega_{H}}{\partial \mu},
\end{equation}
\begin{equation}
p^\|_{MCFL}=-\Omega_{H},
\end{equation}
\begin{equation}
 p^\bot_{MCFL}=-\Omega_{H}+\widetilde{H} \frac{\partial\Omega_{H}}{\partial \widetilde{H}}
\end{equation}
\\
where $\widetilde{H}$ is the in-medium magnetic field and $\Omega_H=\Omega_{MCFL}+B+\widetilde{H}^2/2$ is the thermodynamic potential of the magnetic phase.

\begin{figure}[htb]
  \begin{center}
    \includegraphics[width=0.49\textwidth]{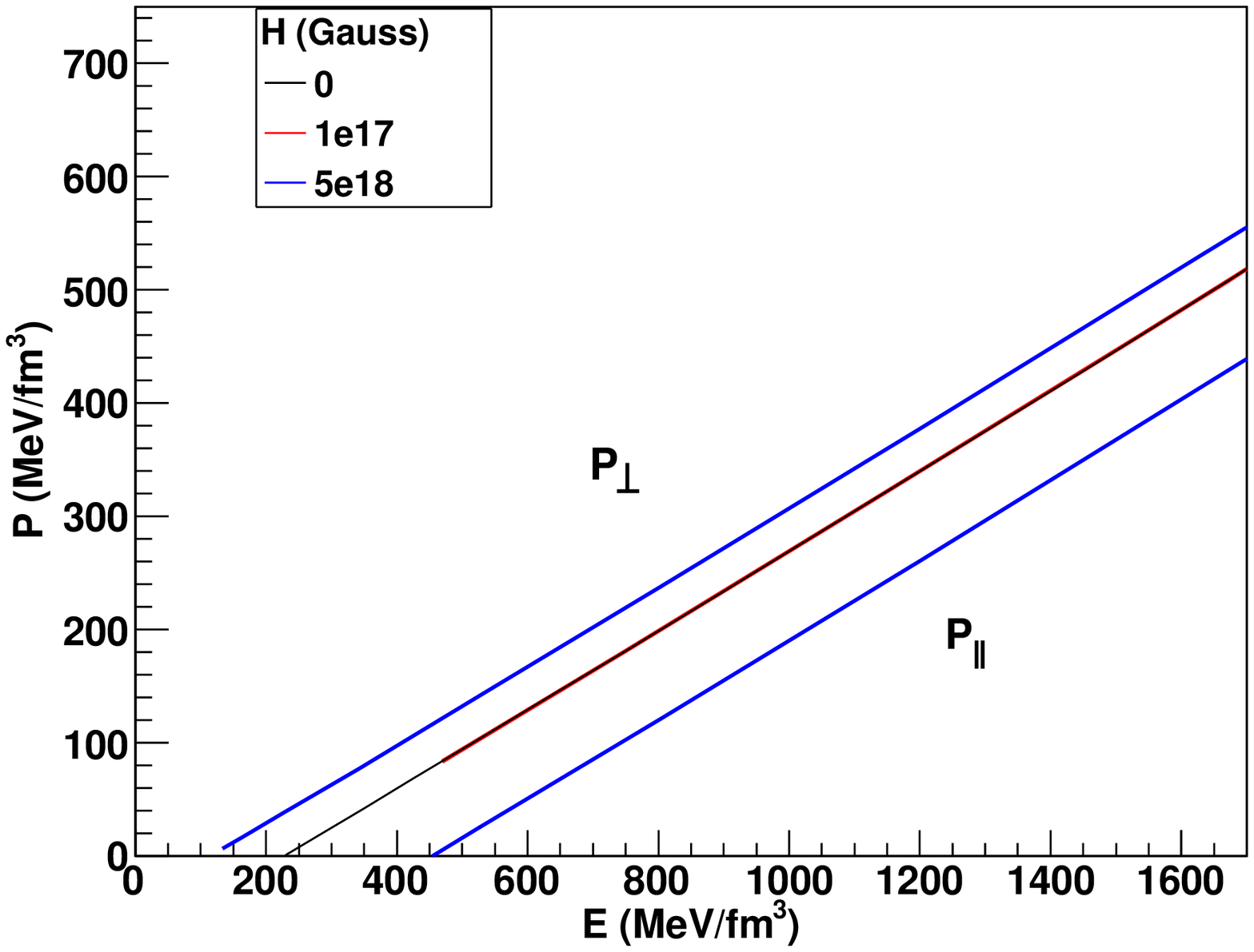}
    \includegraphics[width=0.49\textwidth]{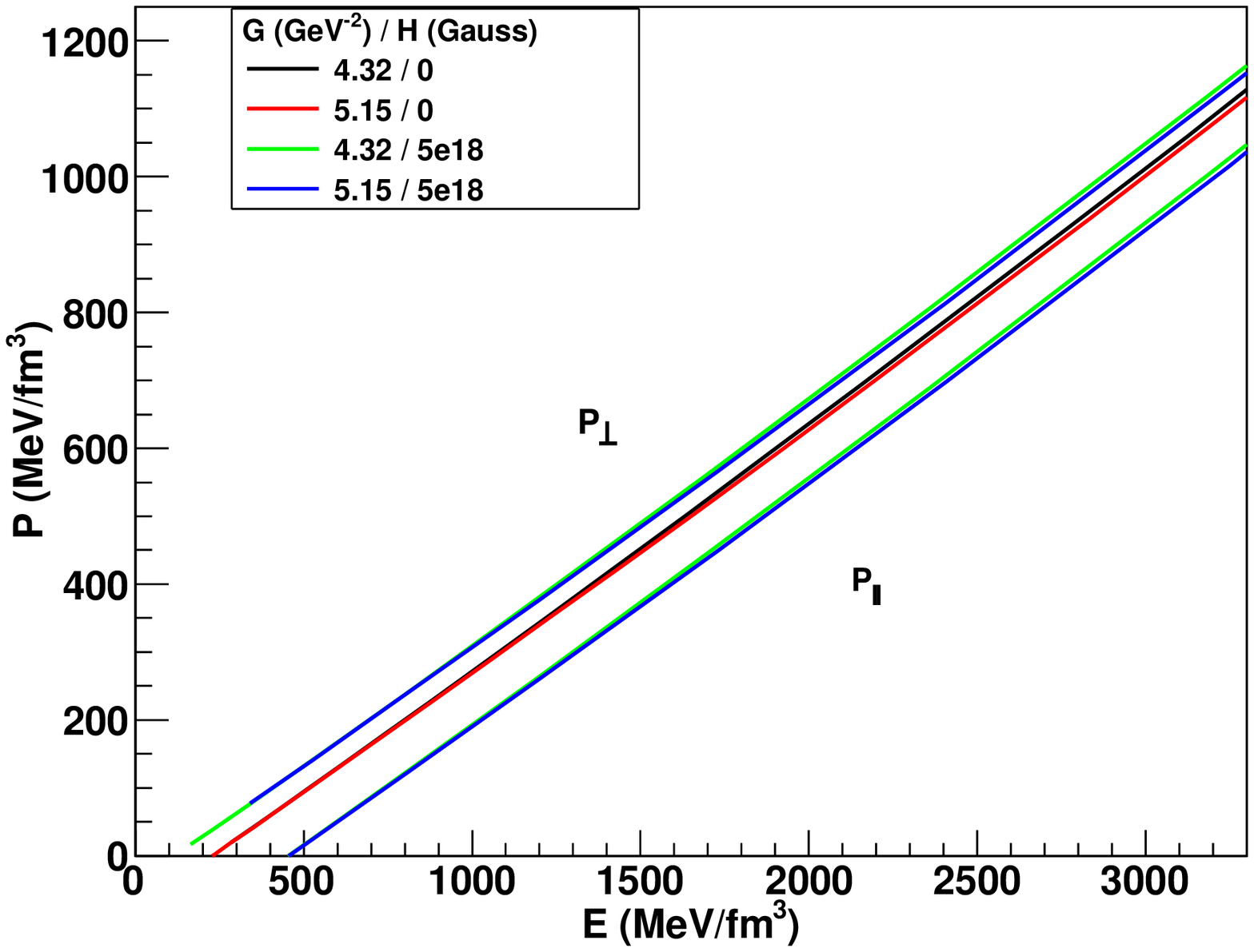}
    \caption{Left panel: EoS for MCFL matter considering parallel and perpendicular pressures for different values of H, as indicated, B = 58 MeV/fm$^3$ and $G$ = 5.15 GeV$^{-2}$. Right panel: Comparison between the EoS for CFL matter for H = 0 and H = $5 \times 10^{18}$ G and $G$ = 4.32 GeV$^{-2}$ and $G$ = 5.15 GeV$^{-2}$ as indicated.} \label{fig1}
  \end{center}
\end{figure}

The EoS remains linear and is substantially modified only at high magnetic fields. Also, for absolute stability to hold in strange matter it is necessary to have a field-dependent bag constant \cite{Paulucci}. In spite of this fact, we kept $B$ constant when calculating the EoS (Fig. \ref{fig1}) since our main goal is to evaluate gravitational bound stars, in which case the star's gravitational field could supply the pressure to compensate the internal one produced by the magnetic field.

These results indicate that from a given value of the magnetic field ($\sim 10^{18}$G), the pressure anisotropy is too large to be dismissed, with the perpendicular pressure giving a ``harder EOS'' than the parallel one. In this way, the Tolman-Oppenheimer-Volkoff (TOV) equations, which apply for stable stars configurations with isotropic EoS, should not be used for obtaining maximum masses for these objects under the influence of strong magnetic fields. For these situations, one must search for a formalism in which the system is considered anisotropic from the beginning making use of cylindrical symmetry.

\section{Conclusions and perspectives}

The EoS is not made harder in the self-consistent approach by using the NJL model with an increase in the value of the gap parameter (corresponding to an increase in $G$). This behavior contradicts the one expected when using the MIT bag model with an approximation to order $\Delta^2$ for the thermodynamical potential. We want to notice that in order to increase the gap parameter at a fixed baryon density in the self-consistent approach, it is needed to strengthen the diquark interaction. Hence, it is found that the system pressure decreases. These combine effects could indicate that the diquark pairs under such strong interaction begins to behave as Bose-Einstein molecules \cite{BEC}. 

Regarding the influence of the magnetic field, on CFL matter it would enforce a new condition (a field dependent vacuum ``bag constant'') for stability. The EoS is substantially modified only at high fields, with a growing anisotropy in the pressure parallel and perpendicular to the field. In this regime, the TOV equations would not be suitable for calculating stability configurations of the star and one would have to work with an anisotropic formalism from the beginning.

We now intend to introduce the diquark repulsion term in the 2SC model to verify the possibility of making the EoS harder. We also intend to calculate hybrid mass-radius sequences and compare with recent measurements for compact stars' masses.

\bigskip
\bigskip
{\bf Acknowledgments:}
The authors wish to acknowledge the financial support received from
Funda\c c\~ao de Amparo \`a Pesquisa do Estado de S\~ao Paulo and
from the CNPq Agency (Brazil). The work of EJF and VI was supported in part by the Office of Nuclear Theory of
the Department of Energy under contract de-sc0002179.

\end{document}

%% file: econfmacros-csqcd3.tex
\def\beq{\begin{equation}}
\def\eeq#1{\label{#1}\end{equation}}
\def\eeqn{\end{equation}}

\def\beqa{\begin{eqnarray}}
\def\eeqa#1{\label{#1}\end{eqnarray}}
\def\eeqan{\end{eqnarray}}

\let\bar=\overbar

\def\Dslash{\not{\hbox{\kern-4pt $D$}}}
\def\dslash{\not{\hbox{\kern-2pt $\del$}}}

\def\msb{{\bar{\ssstyle M \kern -1pt S}}}

\usepackage{fancyhdr,graphicx}